 \documentstyle[12pt]{article}
  \newcommand{\be}{\begin{equation}}
  \newcommand{\ee}{\end{equation}}
  \newcommand{\bdis}{\begin{displaymath}}
  \newcommand{\edis}{\end{displaymath}}
  \newcommand{\eg}{\varepsilon}

    \title{(1+1)-dimensional turbulence.}
 \author{R.Benzi$^{1,2}$, L. Biferale$^{3}$, R. Tripiccione$^4$
   and E. Trovatore$^{5}$}
 
\begin{document}
  \maketitle
  \centerline{$^{1}$  AIPA, Via Po 14, 00100 Roma, Italy.}
\centerline{$^3$ Dipartimento di Fisica, Universit\`{a} 
di Tor Vergata,}
  \centerline{ Via della Ricerca Scientifica 1, I-00133 Roma, Italy.}
  \centerline{$^{4}$ INFN, Sezione di Pisa,}
  \centerline{ S.~Piero a Grado, I-50100 Pisa, Italy.} 
  \centerline{$^{5}$  INFM-Dipartimento di Fisica, Universit\`{a} 
di Cagliari,}
  \centerline{Via Ospedale 72, I-09124, Cagliari, Italy.}
  \date{ }
  \medskip

\footnotetext [2] {On leave on absence from Dipartimento di Fisica, 
			Universit\`{a} di Tor Vergata.}
   
  \begin{abstract}
  \noindent
A class of dynamical models of turbulence living  
on a one-dimensional dyadic-tree structure  is introduced and studied.
The models are obtained as a natural generalization of the popular
GOY shell model of turbulence.  These models are found to be
chaotic and intermittent. They represent the first example
of (1+1)-dimensional dynamical systems possessing non trivial multifractal
properties. The dyadic structure allows 
to study spatial and temporal fluctuations. Energy dissipation
statistics and its scaling properties are studied. 
Refined Kolmogorov Hypothesis is found to hold.

   \end{abstract}

\newpage

\section{Introduction}
 Spatio-temporal intermittency is the most intriguing aspect of a fully
developed three-dimensional turbulent flow. 
Turbulent structures are thought to be
generated by chaotic intermittent energy transfer from 
large to small scales. The cascade is pictorially
described by the Richardson scenario:  large scale 
eddies destabilize and generate  small scale eddies  with
 shorter eddy turnover times. In this way, a 
 hierarchy of fluctuations on smaller and smaller scales and with
 shorter and shorter characteristic times is produced.
 
 Kolmogorov 1941 theory \cite{K} describes the 
 statistics  of velocity differences,
 $\delta_r v$, at  scale $r$ in terms 
of the averaged energy dissipation $\eg$,  neglecting
 completely possible spatio-temporal fluctuations. The velocity
 field statistics is characterized, among others, by the 
 scaling exponents, $\zeta(p)$, of structure functions $S_p(r)$,
  in the inertial range:
\begin{equation}
S_p(r) \equiv <|v(x+r)-v(x)|^p> \equiv <|\delta_r v|^p> \sim 
r^{\zeta(p)}.
\label{eq:zetap}
\end{equation}
In the Kolmogorov description, a simple dimensional argument
leads to the predictions: $$<|\delta_r v|^p> \sim \eg^{p/3} r^{p/3};
\,\,\,\,\,\,\,\,\,\,\,\zeta(p)=p/3.$$
 On the other hand, experiments
show \cite{benzi_ess} that scaling exponents deviate
from the linear behaviour. This departure from linearity 
is the main signature of intermittency and implies
non-gaussian   probability
distribution functions for the velocity differences in the 
inertial range. \\ Intermittency
 also affects energy dissipation statistics. \\ Experiments \cite{pmodel}
show that the energy dissipation defines 
 a multifractal measure on the fluid volume. 
The multifractal measure is characterized
by the  scaling  properties of the coarse-grained
 energy dissipation, $\eg_r$, namely:
\be
<\eg_r^p> \equiv \left< \left({\frac{1}{r^3} \int_{\Lambda(r)} d^3x 
\eg(x)}
\right)^p \right>
 \sim r^{\tau(p)},
\label{eq:taup}
\ee 
where $<\cdot>$ means averaging over all boxes $\Lambda(r)$
 of size $r$ in which the volume occupied by the fluid is partitioned.\\
  The celebrated Refined Kolmogorov Hypothesis
(RKH) \cite{k62} links the statistics of velocity differences in the 
inertial 
range   with the statistics of the coarse-grained  
energy dissipation:
\be
\frac{(\delta_r v)^3}{r} \sim \eg_r
\label{eq:rkh}
\ee
where the symbol $\sim$ means that quantities on  both sides  have
the same scaling properties.

Using  (\ref{eq:rkh}), it is easy to relate the scaling exponents
 $\zeta(p)$ with the
scaling exponents of the energy dissipation:
 \be
\zeta(p)=\tau(p/3) +p/3.
\label{eq:zetatau}
\ee
This relation is the natural consequence of the RKH and it is quite
well satisfied experimentally \cite{rkh_exp}. On the other hand,
 no  satisfactory  theoretical arguments  which predict (\ref{eq:zetatau}) 
have ever been found.

The simplest way to  explain phenomenologically  the presence
of intermittent deviations  consists in 
 describing the energy transfer mechanism in terms of fragmentation
stochastic processes. In these models 
\cite{pmodel,betamodel,logpoisson},
one  introduces a set of  eddies leaving on a dyadic
structure.

Random fragmentation models 
state that the energies contained in eddies  at successive scales
are connected eachother by independent  stochastic variables with 
distribution that 
does not depend on the scale.  For example, in the random-$\beta$-model
we have $\eg_{n} = \beta \eg_{n+1}$,
where  $\eg_n$  denotes 
the energy transfer due to  a typical eddy of size $r_n = 2^{-n} r_0$.
In the random-$\beta$-model also the  
 active volume, $V_n$, occupied by eddies of  size $r_n$,
is supposed to change randomly from scale to scale:
 $V_{n+1}= \beta V_{n}$.
From the definition (\ref{eq:taup})  we have 
$\tau(p) = -\log_2 <\beta^{p-1}>$ and the corresponding expression 
$\zeta(p) = p/3 -\log_2 <\beta^{p/3-1}>$ must be
valid if the RKH holds.
Due to the freedom in the choice of the $\beta$ distribution all these 
models
are able to fit with good accuracy experimental values.
On the other hand, they lack any direct linking with the original
Navier-Stokes (NS) equations.

In this paper, we investigate   a class
of dynamical models which fill the gap between  purely
stochastic fragmentation models and the original  NS
dynamics.  In particular, we define and study a dynamical model
on a (1+1)-dimensional (that is, having space and time dimensions) 
 dyadic tree.
 We decompose  the original velocity field in terms
of fluctuations localized both in Fourier and real spaces.
  One can look at  this model 
 as  an approximation of the original NS
equations in a wavelets basis (see next section).
 Several similar models have been already
investigated in the past by Grossmann and coworkers
\cite{grossmann1,grossmann2}.
At  difference from  models studied in  \cite{grossmann1,grossmann2},
 we do not impose all the original geometrical constraints
of the NS equations. For example, we do  not have
neither divergenceless velocity fields nor three-dimensional structure
of the real space. In this way we can embed the model in a 
one-dimensional
real space and simplify enormously the structure of the nonlinear terms.

The most important advantage is that  we can 
increase the total number of resolved
scales (and therefore we can reach high Reynolds numbers),
paying the  price of having a model which is not  exactly
derivable from the original NS equations. 
Similarities and differences between our
work and the works of Grossmann and coworkers can be found in the 
conclusions.

The main result of this paper is to present for the first
time a one-dimensional dynamical system sharing some of the most 
important
properties with a real three-dimensional turbulent flow. In particular,
we analyze  
  structure functions intermittency and the energy dissipation spatial
distribution. RKH is
found to be 
remarkably well satisfied.  These models are the natural ground
where testing new developments of large-dimensional dynamical-systems
 theory and new approaches to turbulent (i.e., multi-scale systems) flows. 
 
As we discuss in the following, 
numerical simulations need  state-of-the-art multi-processor 
computers.

The paper is organised as follows. In Sec.2, we discuss how to
 jump from zero-dimensional shell models to
one-dimensional spatially-resolved tree models; 
the tree model is introduced in Sec.3; in Sec.4, numerical results
on structure functions
intermittency, viscous scales dynamics
and energy dissipation field statistics are presented;
conclusions follow in Sec.5. 

%%%%%%%%%%%%%%%%%%%%%%%%%%%%%%%
\section{From chains to trees}

In order to understand turbulent energy transfer dynamics
 and related intermittent
effects, dynamical deterministic models have been proposed. 
Among these models, shell models have recently attracted
the interests of many researchers (see \cite{KA} for a popular 
introduction).
The basic idea in such models is to retain only a few variables 
as representative  of  an entire {\it shell} of Fourier 
amplitudes. The 
nonlinear structure of NS equations is preserved,
 but all three-dimensional properties of the original embedding
space are lost.

The most popular shell model is  the Gledzer-Ohkitani-Yamada 
(GOY) model (\cite{G}-\cite{BKLMW}).

The GOY model can be seen as a severe truncation of the 
NS equations. Dynamical variables are described
by  a single  
 complex mode $u_n$ for each   
shell of wave numbers $k$ such as  $k_n < k < k_{n+1}$,
with $k_n = \lambda^n $ 
and  $\lambda$
being an arbitrary scale parameter ($\lambda>1$), usually 
taken equal to $2$. 
In the equations, only quasi-local couplings between nearest and next 
to nearest 
neighbor shells are kept.

Recently, a new class of shell models based upon the helical
 decomposition of NS equations \cite{W}
has been suggested \cite{BK} and studied \cite{BBKT}.  
In this way, it is possible to obtain a second non-positive defined 
invariant 
closer to the   NS helicity. 
The models, which have now two complex variables per shell 
($u_n^+$ and $u_n^-$, 
transporting
positive and negative helicity respectively) are simple  
generalizations of the GOY model. From now on we will concentrate
only on one of them (see below), which has been previously shown to 
share 
many properties  with true turbulent three-dimensional flows \cite{BBKT}.

In particular, in \cite{BBKT} it has been shown that this model
has the {\it same} degree of intermittency found 
 experimentally  in fully developed turbulence, if  
 the set of free parameters
is chosen in order to
conserve energy and helicity in the inviscid and unforced 
limit.  

Shell models can be thought of as  field 
problems in zero spacial dimension: their obvious limitation consists
in treating all degrees of freedom in a Fourier shell at once; the physical 
object they describe are coherent planar waves, 
filling the whole volume of the fluid.

Real turbulence consists of localized eddies of all sizes that interact, 
merge and subdivide locally: the physical picture is that of a 
large eddy  which decays into smaller eddies. 
The number of degrees of freedom in such a field problem in
$d$ dimensions grows with the wave number as $N(k) \sim k^d$
($d=0$ in  shell models).
The first step in reproducing 
this kind of hierarchical structure is to transform  a  {\it chain}-model 
 into a {\it tree}-model with $d=1$.
This is achieved by letting grow the number of degrees of freedom
with the shell index $n$ as $2^n$.

As in the original shell models, this tree model
must be  in some sense reminiscent of the  NS equations.
It can be regarded as 
describing the evolution of the coefficients of an orthonormal wavelets 
expansion of a one-dimensional projection of the 
velocity field $v(x,t)$:

\begin{equation}
v(x,t)=\sum_{n,j} \hat{v}_{n,j}(t) \psi_{n,j}(x).
\label{eq:wave}
\end{equation}

Here $\psi_{n,j}(x)$ are a complete orthonormal set of wavelets 
generated from
an analyzing wavelet $\psi_{0,0}(x)$ by discrete translations and 
dilations:

\begin{equation}
 \psi_{n,j}(x)= 2^{n/2} \psi_{0,0}(2^n x-j).
\label{eq:wave1}
\end{equation}

In principle, it is possible to plug the three-dimensional generalization 
of (\ref{eq:wave1}) in the NS equations and deriving an 
exact dynamical evolution for the wavelets coefficients \cite{nakano}.
For the sake of simplicity, we will be interested  to some approximated 
one-dimensional truncation of the wavelets resolved dynamics. In 
particular,
we can construct dynamical equations of the type of shell models
for our tree variables, viewed  as the analogues 
of the expansion coefficients $\hat{v}_{n,j}$.
Similar wavelets decomposition, but with purely stochastic coefficients
$\hat{v}_{n,j}$ have been used for defining synthetic multi-affine 
signals
\cite{BBCPVV}.

In fig.1, we pictorially show our tree structure, covering the 
one-dimensional interval $[0,\Lambda_T]$.
Each dynamical variable $\hat{v}_{n,j}$ is represented by a box of length
$l_n=2^{-n}$, occupying the region $\Lambda_j(n)$ ranging from $(j-1)l_n$
to $j l_n$.
At each scale $n$ there are $2^{n-1}$ boxes, covering a total length 
$ \Lambda_T=2^{n-1} l_n=1/2 $.

For the sake of convenience we define the tree model in terms of {\it
density} variables,
$u_{n,j}$ (depicted as balls in figs.2,3), which would cor\-re\-spond to  
$\hat{u}_{n,j}= 2^{n/2} \hat{v}_{n,j}$ in a wavelets expansion.

In this notation,  
$| u_{n,j} |^2$ 
represents the energy density in a flow structure of length
$l_n=2^{-n}$ and spatially labeled by the index  $j$.

%%%%%%%%%%%%%%%%%%%%%%%%%%%%%%%
\section{The tree model}

We have chosen the helical shell model studied in \cite{BBKT}
as the starting point for the construction
of our hierarchical structure.
The dynamical equations of this shell model are the following:    

\begin{equation}
\dot{u}^+_n=i k_n (
a \,\, u^{+}_{n+1} u^{-}_{n+2}+
b \,\, u^{+}_{n-1} u^{-}_{n+1}+ 
c \,\, u^{-}_{n-2} u^{-}_{n-1}+
)^*
-\nu k^2_n u^+_n +\delta_{n,n_0} F^+,\\
\label{eq:shells}
\end{equation}

and the same holds, with all helicities reversed, for $\dot{u}^-_n$.
Here, $n=1,...,N$, where $N$ is the total
number of shells, $\nu$ is the viscosity, $F^+$ the external forcing acting
on a large-scale shell $n_0$ and $a$,$b$,$c$ are three parameters, 
which 
are determined by imposing
 conservation of energy
and  helicity in the inviscid and unforced limit:

\begin{equation}\begin{array}{ll}
 &dE/dt =d/dt \,\,(\sum_n \vert u^+_n \vert^2 + 
\vert u^-_n \vert^2) = 0,\\
 &dH/dt =d/dt \,\,(\sum_n k_n 
(\vert u^+_n \vert^2 - \vert u^-_n \vert^2))= 0.
\end{array}
\label{eq:shells1}
\end{equation}

The statistical properties of this model have been studied in 
\cite{BBKT}: 
the system  turned out to have
an intermittent energy transfer 
 very similar to what one can find in the original NS equations.
 
Hereafter we will fix 
the intershell separation $\lambda=2$.
For this particular (and standard) choice, we must fix:
$a=1$, $b=-5/12$ and $c=-1/24$.

We introduce a  spatial degree of freedom in the system by
 using the notation $u^{\pm}_{n ,j}$ to indicate the complex 
helical variable on scale $n$ and spatial position labeled by the index 
$j$.
For a given shell $n$, the index $j$ can vary from $1$ to $2^{n-1}$.

In this tree structure, each variable $u_{n ,j}$  
 continues to interact with the nearest and next nearest
levels, as in equation (\ref{eq:shells}); however, a variety
 of possibilities is now opened by the presence of many 
 horizontal degrees of 
freedom localized on each shell.
 
 The simplest choice is depicted in fig.2, where a portion of 
 the tree structure is showed and the evolving in time variable, 
  $u_{n ,j}$,
 is represented by a black ball.
 In the figure, solid lines connect interacting balls (variables); 
 in particular,
 several solid lines connecting one ball
 on a level to a group of balls, all belonging to another level,  
 mean a sum  over the corresponding 
 variables in the equation.
 For simplicity, we will omit to write down explicitly 
these sums in the equations, 
 avoiding to specify
 the second (spatial) index $j$ of the variables when is univocally
 determined from figures or in case of the implicit above
mentioned  sums.
 Thus, by writing the dynamical tree equations as follows:
 
\begin{equation}
\dot{u}^+_{n,j}=i k_n (
a/4 \,\, u^{+}_{n+1} u^{-}_{n+2}+
b/2 \,\, u^{+}_{n-1} u^{-}_{n+1}+ 
c \,\, u^{-}_{n-2} u^{-}_{n-1}
)^*
-\nu k^2_n u^+_{n,j} +\delta_{n,n_0} F^+,\\
\label{eq:tree_delta0}
\end{equation}
  
we mean, for example, in the first term of the RHS of (\ref{eq:tree_delta0}) 
 (see fig.2a):
\begin{equation}
a/4 \,\, [ 
 u^{+}_{n+1 ,2j-1} (u^{-}_{n+2 ,4j-3} + u^{-}_{n+2 ,4j-2})+
 u^{+}_{n+1 ,2j} (u^{-}_{n+2 ,4j-1} + u^{-}_{n+2 ,4j}) ].
\label{eq:exam}
\end{equation}
 
Here again the same holds, with all helicities reversed,
 for $\dot{u}^-_{n ,j}$.
 The numerical values of $a$, $b$ and $c$ are the same of the original
helical shell.
 In the unforced and inviscid limit our system conserves the total energy
and helicity, namely:

\begin{equation}\begin{array}{ll}
&dE/dt =d/dt \,\,(\sum_{n,j} 2^{-n}  
| u^+_{n,j} |^2 + |u^-_{n,j} |^2)= 0, \\
&dH/dt =d/dt \,\,(\sum_{n,j} 2^{-n} k_n 
| u^+_{n,j} |^2 - |u^-_{n,j} |^2)= 0.
\end{array}
\label{eq:inv}
\end{equation}

Let us remind that our tree variables can be roughly viewed as the 
one-dimensional model counterparts of the amplitudes, $\hat{u}_{n j}=
2^{n/2} \hat{v}_{n j }$, in a wavelets expansion of the velocity
field (see eq.(\ref{eq:wave1})). 

This analogy must be carried only on qualitative grounds, in particular 
if one considers that, in the wavelets basis, the Laplacian of the 
viscous term is not diagonal and that we have
drastically reduced the possible interactions among variables.
However, in a wavelets expansion, we expect
the nonlinear and dissipation terms to be large only if  the 
involved scales 
 are of similar sizes: this corresponds to the physical 
property that eddies in a fluid mainly 
transfer energy to eddies of similar scale.
Also interaction among the position indices can be expected to be 
localized,
with the precise degree of localization depending on 
the details of the wavelets basis. 

Besides the simple implementation (\ref{eq:tree_delta0}),
one can build up more complex equations,
simply turning on new interactions among variables (always
conserving energy and helicity).
The guide line in restricting the possible choices can
be phenomenologically motivated by requiring 
a certain degree of locality, both in Fourier and real spaces.

As an example, an enlarged set of interactions is depicted in fig.3:
in particular two contributes reminiscent of the 
Desnyansky-Novikov (DN) shell model \cite{DN} have been added 
(see figs.3d,3e). 
Structurally new interactions are those depicted in 
fig.3f, where only horizontal couplings are considered.

By putting all together we have:

\begin{equation}\begin{array}{llll}
\dot{u}^+_{n,j}=&i k_n [
a/4 \,\, u^{+}_{n+1} u^{-}_{n+2}+
b/2 \,\, u^{+}_{n-1} u^{-}_{n+1}+
c \,\, u^{-}_{n-2} u^{-}_{n-1}+\\
&d \,\, (- u^{-}_{n+1} u^{+}_{n+1}- u^{+}_{n+1} u^{-}_{n+1}+
e_1 \,\, u^{-}_{n} u^{+}_{n-1}+
e_2 \,\, u^{-}_{n} u^{-}_{n-1} )+ \\
&f \,\, (  u^{+}_{n ,j+1} u^{-}_{n ,j+2}-
  u^{+}_{n ,j-1} u^{-}_{n ,j+1}-
  u^{-}_{n ,j-1} u^{+}_{n ,j+1}
+ u^{-}_{n ,j-2} u^{+}_{n ,j-1})
]^*\\
&-\nu k^2_n u^+_{n,j} +\delta_{n,n_0} F^+,
\end{array}
\label{eq:tree_delta1}
\end{equation}

where the spatial index is shown only in the terms 
corresponding
to the horizontal interactions of fig.3f, i.e., in the third row of 
eq.(\ref{eq:tree_delta1}).
In considering these particular terms, the tree must be viewed as leaving 
on the surface of a  cone, with the last 
position $j$ on each level connected to the first one on the same level.

All coefficients in the nonlinear terms of (\ref{eq:tree_delta1})
 are chosen in order to conserve the total 
energy and helicity (\ref{eq:inv}) of the system.
Each of the three interactions types (GOY, DN and horizontal) 
conserves
separately E and H: this leaves some freedom only in
 the choice of the two relative weights
 $d$ and $f$ in (\ref{eq:tree_delta1}). 
  We fixed $d=f=1$; while the coefficients $a, b, c$ 
  are the same as before and  $e_1=3/4$ and $e_2=1/4$.

In our simulations we have considered both the  simple version  
of the hierarchical system described by 
(\ref{eq:tree_delta0}) and by  fig.2,
which we will call version A,
 and the more complete described by  (\ref{eq:tree_delta1}) and by 
fig.3, denoted 
hereafter as version B.

\section{Numerical results}

\subsection{Numerical implementation}

In our simulations we have considered a total number of levels 
$N=16$. 
The  total number of sites forming the tree is then $N_T = 2^N-
1=65535$,
each one  described in terms of two complex variables.

When dealing with this tree structure, the computer effort needed for  
numerical simulations
 increases enormously with respect to the original chain model: 
in order to collect a reliable statistics 
 a computing time  $O(10^4)$ longer 
is now needed.

For this reason, we have implemented our numerical experiments on
the APE-100 machine \cite{ape}.

APE-100 is a Single Instruction Multiple Data (SIMD) Parallel Architecture,
based on a three dimensional cubic mesh of nodes with periodic boundary 
conditions, each node being connected to its 6 neighbors.
The particular version we used is a 512-node configuration, with
a floating point performance of 50 Mflops peak speed
on a single processor.
Each processing node contains a floating point 32 bit real 
arithmetics processor and a 4 Mbyte local memory.

We used the parallelism of this machine in the simplest
way: its 512 processors integrate simultaneously the same equations,
 starting from different initial conditions.
A good statistics is then obtained by time-averaging on each single
processor and ensemble-averaging over all processors.

In both cases A and B, we put the following parameters values
into the dynamical equations:

\begin{equation}\begin{array}{lllll}
& N=16, \\
&\nu=2.0 \,\,10^{-5},\\
&F^{\pm}=(5.0 \,\,10^{-3},5.0 \,\,10^{-3}),\\
&n_0=1,\\
&k_0=6.25 \,\,10^{-2}.
\end{array}
\label{eq:param2}
\end{equation}
  
We used the Runge-Kutta fourth-order method in integrating the 
system,
with a time step varying from $10^{-3}$ to $5.0\cdot10^{-4}$ time 
units,
 depending on case A or B.
In both cases we integrated the model for a total number of steps
of the order of $10^5$ on each processor of the parallel machine;
considering all processors,
we were able to collect a total statistics of 
 500 eddy-turnover times, using few days of computing time.

\subsection{Syncronization}

Before considering the statistical properties of the system,
 let us mention a dynamical effect which emerged 
immediately 
in numerical integrations, that
is, synchronization on the last (dissipative) levels.

Even starting to integrate from an initial
configuration having different values on different sites of each
tree level, dynamical variables on nearby locations in the last shells
(i.e., in the dissipative range) synchronize after  relatively few time steps.

The mechanism underlying this phenomenon can be easily understood if
one considers what follows:
at the viscous cutoff (which corresponds in our case to
 $ n_d \sim 11$), 
the ratio between variables belonging
to two successive scales $|u_n|/|u_{n-1}|$ 
undergoes an abrupt decrease, when passing from 
the inertial range ($n \le n_d$)
to the viscous range ($n > n_d$).
Then, in the dynamical equation for $n=n_d$,
interactions  involving the upper levels become largely predominant  
with respect to the others: this naturally leads to a similar
behaviour of  variables sitting on 
two close locations, whose dynamics turns out to be governed
by essentially the same equation.
This synchronization is then transmitted 
and spread out into  lower levels,
in which nearby variables become equal in groups of 4, 8, 16, .... and so 
on,
as the shell index $n$ increases. 

In both cases A and B, 
this synchronization does not affect the inertial shells. 
In the version A of the model
this process starts already from $n^*=11$. In case B
 the situation changes: the horizontal interactions  
 are quite effective in  breaking the synchronization on two close 
sites nearby the dissipative cutoff 
and  synchronization effects are thus shifted
towards the very last levels ($n=$15 and 16, in our case).

\subsection{The energy dissipation field}

The first step in constructing the energy dissipation field
of our model is to consider the following function:

\begin{equation}
\eta_{n,j} = \nu k^2_n (|u^+_{n,j}|^2 + |u^-_{n,j}|^2),
\label{eq:eta}
\end{equation}

which represents the energy dissipation {\it density} 
in the structure  covering the region $\Lambda_j(n)$
of length $2^{-n}$, centered in the spatial
site labeled by $j$.
These structures are represented by boxes in fig.1.

The total energy dissipation density,
$\epsilon = (1/\Lambda_T) \int_{\Lambda_T} \epsilon(x) \, dx $,
where $\Lambda_T$ is the total space length, 
is, by definition,  the sum of all these contributes
(sum over boxes at all scales in fig.1):

\begin{equation}
\epsilon = \sum_{n,j} 2^{-n} \eta_{n,j}.
\label{eq:e1}
\end{equation}

On the other hand, in order to study the scaling properties 
of the energy dissipation field, one has to disentangle in 
 $\epsilon$ the  contributions coming 
from the coarse-grained energy dissipation field $\epsilon_r$, as   
 defined in Sec.1,  
 eq.(\ref{eq:taup}).  
In our formulation, we can then rewrite:

\begin{equation}
\epsilon = \frac{1}{\Lambda_T} \int_{\Lambda_T} \epsilon(x) \, dx =
\frac{1}{2^{n-1}} \sum_{j=1}^{2^{n-1}} 
\left( {\frac{1}{2^{-n}} \int_{\Lambda_j(n)} \epsilon(x) \, dx }\right)=
\frac{1}{2^{n-1}} \sum_{j=1}^{2^{n-1}}  \epsilon_{n,j},
\label{eq:e2}
\end{equation}

where the last expression
 is independent of $n$ and 
 the $\epsilon_{n,j}$'s are the coarse-grained energy
 dissipation densities,
obtained as averages over spacial regions of length $2^{-n}$.
Note that the average density $\epsilon_{n,j}$ over $\Lambda_j(n)$
does not coincide simply with the density $\eta_{n,j}$ of the
structure living in $\Lambda_j(n)$, namely:

\begin{equation}
\epsilon_{n,j} = \eta_{n,j} + \sum_{m < n} \eta_{m,k(m)} +
\sum_{m > n} <\eta_{m,k(m)}>_{I(m)}.
\label{eq:e3}
\end{equation}

Here, in the second (third) term of the RHS we take into account
density contributions coming from larger (smaller) scale structures
(as an example, all regions contributing to the definition
of $\epsilon_{n,j}$ are represented as shadowed boxes in fig.1).
The index $k(m)$ in the second term of RHS
 labels the  location of larger scale
structures containing the region  $\Lambda_j(n)$ 
under consideration (shadowed boxes with $m < n$ in fig.1). In the third term, 
an average  is performed over  
 $k(m) \in I(m)$, where $I(m)$ labels the set of structures 
  contained in  $\Lambda_j(n)$, for any $m > n$
  (in fig.1, $I(m)$ labels the two boxes at $n+1$,
  the four boxes at $n+2$, and so on).

The best spatially resolved energy dissipation field is for $n=N$:

\begin{equation}
\epsilon_{N,j} = \sum_{m \le N} \eta_{m,k(m)};
 \,\,\,\,\,\,\,\,\,\,\,\,\,\,\,\,\,\,\,\,\,\,\,\,\,\,\, j=1,...,2^{N-1}.
\label{eq:e(x)}
\end{equation}

In fig.4,  the instantaneous values assumed by
$\epsilon_{N,j}$ in the $N_T/2=32768$ locations of the last 
level are showed.
The chaotic, intermittent character of this spatial
signal is evident.

\subsection{Scaling laws and the 
Refined Kolmogorov Hypo\-the\-sis}
 
Performing long-time numerical integrations of 
eqs.(\ref{eq:tree_delta0},\ref{eq:tree_delta1}), we have 
studied the statistical properties of both 
 versions A and B of our tree model.
We have investigated the scaling properties of 
velocity field structure
functions, $S_p(n)$, and of coarse-grained energy dissipation moments, 
$D_p(n)$.

These moments 
have been evaluated by time,space and ensemble--averaging 
our variables over the total integration time,
the spatial locations and the processors of the APE machine:

\begin{equation}\begin{array}{lll}
S_p(n) &\equiv 
\sum_{m,t,j} {\frac{1}{M \, T \, 2^n} 
\left( {\sqrt{|u^+_{n,j}(t,m)|^2+
|u^-_{n,j}(t,m)|^2}}\right)^ {\; p}},\\
& \\
D_p(n) &\equiv 
\sum_{m,t,j} {\frac{1}{M \, T \, 2^n} {\epsilon}^  p_{n,j}(t,m)},
\end{array}
\label{eq:average}
\end{equation}

 where $m$ is the processor index (varying from 1 to $M=512$)
 and $t$ is the time step index.
 
 These moments are expected to follow, in the inertial
 range, the scaling laws (see eqs.(\ref{eq:zetap},\ref{eq:taup})): 

\begin{equation}\begin{array}{ll}
 S_p(n) &= k_n^{- \zeta (p)}, \\
 D_p(n)  &= k_n^{- \tau (p)},
\end{array}
\label{eq:explaw}
\end{equation}

with the two scaling exponents related together by the RKH relation 
(\ref{eq:zetatau}), eventually.
  
 In  fig.5, the log-log plot  of the sixth order
 structure function $S_6(n)$ against  $k_n$ is presented for the 
 two versions A and B:
 the two different slopes in the inertial range  indicate
 a degree of intermittency strongly dependent on how the 
 variables of the tree interact.
 
 Let us notice that in both 
 versions A and B, the interaction range in Fourier space
 is the same
 (a typical eddy at scale 
 $n$ interacts always with the nearest and nextnearest scales
 $n-1$, $n+1$, $n-2$ and $n+2$); 
 nevertheless, in the two cases the interaction range
 in physical space  is different and a different number of connections 
 between  scales is considered. 
 This seems to be of primary importance in determining the intermittency
 degree of the system.
 
 In order to obtain a quantitative measure of the
 structure functions scaling exponents,
 we performed a fit over the inertial range shells,
 using Extended Self Similarity (ESS) \cite{benzi_ess}:
 in fig.6, the logarithm of the sixth order moment $S_6(n)$
 against the logarithm of the third order moment $S_3(n)$ is
 plotted, again for both versions A and B.
 The resulting exponents, $\zeta(p)/\zeta(3)$, obtained for all the 
 moments $p=1,...,8$, are reported in table~1 and
 compared with the K41 nonintermittent prediction $p/3$.

 We found
 a similar scaling behaviour
 for the energy dissipation moments: as an example,
  we show in fig.7 the
 log-log plot of $D_2(n)$ against $k_n$; again,
 the reduced steepness in case B indicates a less intermittent 
 behaviour, with $\tau(p)$ closer to the zero value. 

The subsequent step was to test the RKH
for our tree model: to improve the fitting procedure,
we applied the ESS method  to the RKH relation \cite{benzi_ess}:

\begin{equation}
S_p(n) = (S_3(n))^{p/3} D_{p/3}(n).
\end{equation} 

Plotting the LHS of this equation against the RHS in a 
log-log plot, one can directly test the validity of
relation  (\ref{eq:zetatau}) in the inertial region:
in fig.8  three cases, with $p=4$, 5, 6, 
 are reported, for the version A; the same, but for version B, are in fig.9.

In both versions A and B and for all the moments considered,
our fits confirm that the RKH is 
well satisfied, being all the slopes  in the 
inertial region equal to 1 within a few percent. 

The validity of the RKH  confirms
that dissipative quantities have non trivial influences on inertial
range dynamics (and viceversa). The  tree-like structure 
imposed on the velocity fluctuations does not necessary implies
 that the energy dissipation can be described in terms of
fragmentation processes. In order to test the scale-organization
of the energy structures, ultrametric-sensitive observables should
be studied. 

\section{Conclusions} 
 
 A new class of dynamical models in one spatial dimension which 
show spatial and temporal chaos (fully developed turbulence) has been
introduced and numerically studied.\\
The model originates from a wavelets-like decomposition of a one-dimensional
 cut of a turbulent velocity field. Mainly local interactions in space and 
in scale  are retained. \\
Structure functions scaling shows typical multifractal behaviour of a 
real turbulent flow. Also  spatial fluctuations of the energy dissipation
are in qualitative and quantitative agreement with the corresponding 
measured quantities in  a real fluid. Refined Kolmogorov Hypothesis 
linking the statistics of structure functions to the coarse-grained 
energy dissipation is found to hold within a few per cent.\\
This kind of models open the possibility to investigate numerically
many open questions of  turbulence. Among them, work is in 
progress for detecting possible ultrametric structure in the energy 
cascade mechanism \cite{menevau}. \\
The presence of a real space allows also to test in a quantitative way
some of the most popular eddy-viscosity models used for simulating 
small-scale activity in Large-Eddy-Simulations.
 
Our results show that intermittency 
strongly depends on the degree of tree connectiveness.
For example, by passing from version A to version B one observes
a decreasing in the 
 deviations $\delta\zeta(p)=|\zeta(p)-p/3|$ from Kolmogorov scaling
(from $\sim  30\%$ in case A to $\sim 10\%$ in case B): this
is probably due to the presence of horizontal couplings in version B 
which allow a better efficiency in the energy exchanges. 
\noindent
Increasing the number of triads-couplings should enlarge
 the number of possible downward paths followed by the energy cascade.
 
This tendency toward a less intermittent regime, by increasing the number of 
triad interactions, may seem in contrast with the observed
intermittency in the original Navier-Stokes equations (which 
have all possible interactions switched on). This
contradiction is only apparent: divergenceless character
of the original NS field, added to complicated phase-coherence effects,
can   very easily introduce different dynamical weights in the possible
 triad interactions  leading to a situation
where only a few of them govern the global dynamical evolution.
For example, Grossmann and coworkers showed, by performing 
suitable truncation of NS equations,
that intermittency depends on the typical  degree
of locality  in Fourier space of the survived triad interactions;
very similar results have also been found in shell models 
at varying the inter-shell ratio $\lambda$ \cite{bbt_prl}; the question
of finding universal observables in hydrodynamics-like intermittent
systems remains of primary interest \cite{bbt_prl,sw,glr}.

\vskip0.5cm
\noindent
ACKNOWLEDGEMENTS

Interesting discussions with E.~Aurell and D.~Lohse are kindly acknowledged.
During the preparation of this paper we have received a preprint by E. 
Aurell, E. Dormy and P. Frick on  a similar generalization of shell 
models for two dimensional turbulence. They give slightly
different explanation 
for synchronization of small scales dynamics.

%%%%%%%%%%%%%%%%%%%%%%%%%%%%%%%

%%%%%%%%%%%%%%%%%%%%%%%%%%%%%%
\newpage

\centerline{FIGURE CAPTIONS}

\begin{itemize}

\item Figure 1: A picture of the hierarchical system, covering the 
		one-dimen\-sio\-nal interval $[0,\Lambda_T]$.

\item Figure 2: Pictorial representation of nonlinear interactions
		of eq.(\ref{eq:tree_delta0}).

\item Figure 3: Pictorial representation of nonlinear interactions
		of eq.(\ref{eq:tree_delta1}).
		
\item Figure 4: Instantaneous configuration of the coarse-grained 
		energy dissipation
		density field, $\epsilon_{N,j}$, over the last level sites
		(version B).
		
\item Figure 5:	Log-log plot of the sixth order velocity field structure 
		function, $S_6(n)$, against the wave number $k_n$, 
		for versions
		A (white circles) and B  (black diamonds).
		 
\item Figure 6: Log-log plot of the sixth order velocity field structure 
		function, $S_6(n)$, against the third order velocity field
		 structure 
		function, $S_3(n)$, for  versions
		A (white circles) and B  (black diamonds). Straight 
		lines correspond to a  linear fit in the inertial range. The
		corresponding slopes are reported in table 1.
		
\item Figure 7: Log-log plot of the second moment, $D_2(n)$, of
		energy dissipation density against the wave number $k_n$,
		for  versions
		A (white circles) and B  (black diamonds).
		
\item Figure 8: Log-log plot of the velocity field structure 
		functions, $S_p(n)$ ($p=4,5,6$), 
		against $((S_3(n))^{p/3} \,D_{p/3}(n))$ in the
		inertial range,
		for version A.
		Straight lines correspond to a  linear fit: 
		all cases 
		are  compatible with the RKH.

\item Figure 9: The same as in fig.8, but for version B.

\end{itemize}

\centerline{TABLE CAPTION}

\begin{itemize}

\item Table 1: The scaling exponents ratios, $\zeta(p) / \zeta(3)$ 
		($p=1,...,8$), resulting from the  ESS fitting procedure.
		The values obtained for  
		 versions A and B  are reported in the 
		 second and third column, respectively.
		 In the last column there are the K41 values.

\end{itemize}

\newpage
%%%%%%%%%%%%%%%%%%%%%%%%%%%%%%%
\begin{table}
\begin{center}
\begin{tabular}{|c|c|c|c|}
\hline 
%\cline{2-4}
$p$ & $A$ & $B$ & $K41$\\
\hline
$1$ & $0.41 \pm 0.01$ & $0.348 \pm 0.005$ & 0.333\\
$2$ & $0.74 \pm 0.01$ & $0.682 \pm 0.005$& 0.667\\
$3$ & $1$  & $1$ & 1 \\
$4$ & $1.21 \pm 0.01$ & $1.303 \pm 0.006$ & 1.333\\
$5$ & $1.36 \pm 0.02$ & $1.59 \pm 0.01$ & 1.667\\
$6$ & $1.48 \pm 0.03$ & $1.86 \pm 0.02$ & 2\\
$7$ & $1.55 \pm 0.05$ & $2.12 \pm 0.03$ & 2.333\\
$8$ & $1.60 \pm 0.07$ & $2.35 \pm 0.03$ & 2.667\\
\hline
\end{tabular}
\end{center}
\label{tab}
\end{table}

%%%%%%%%%%%%%%%%%%%%%%%%%%%%%%
  
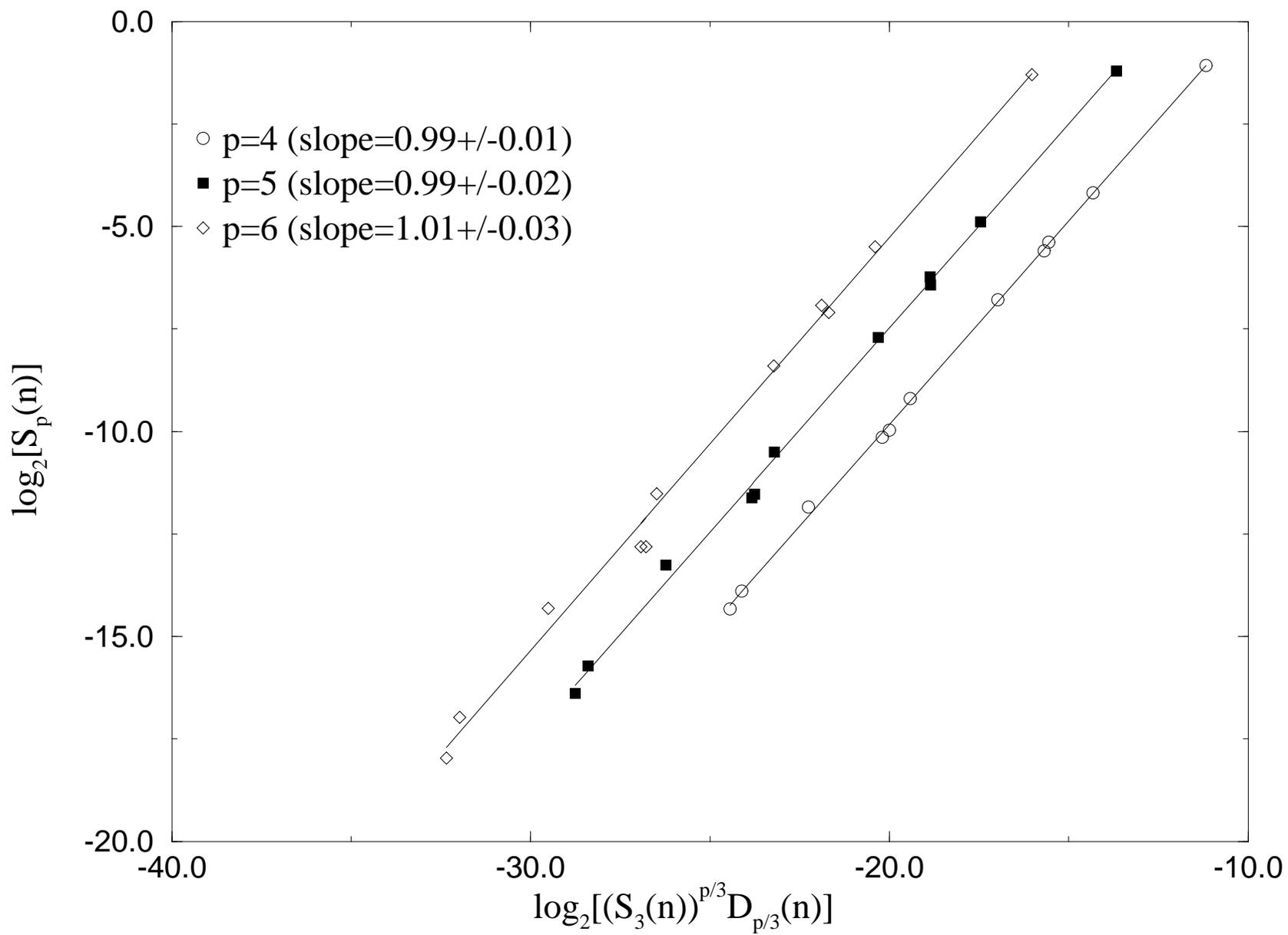
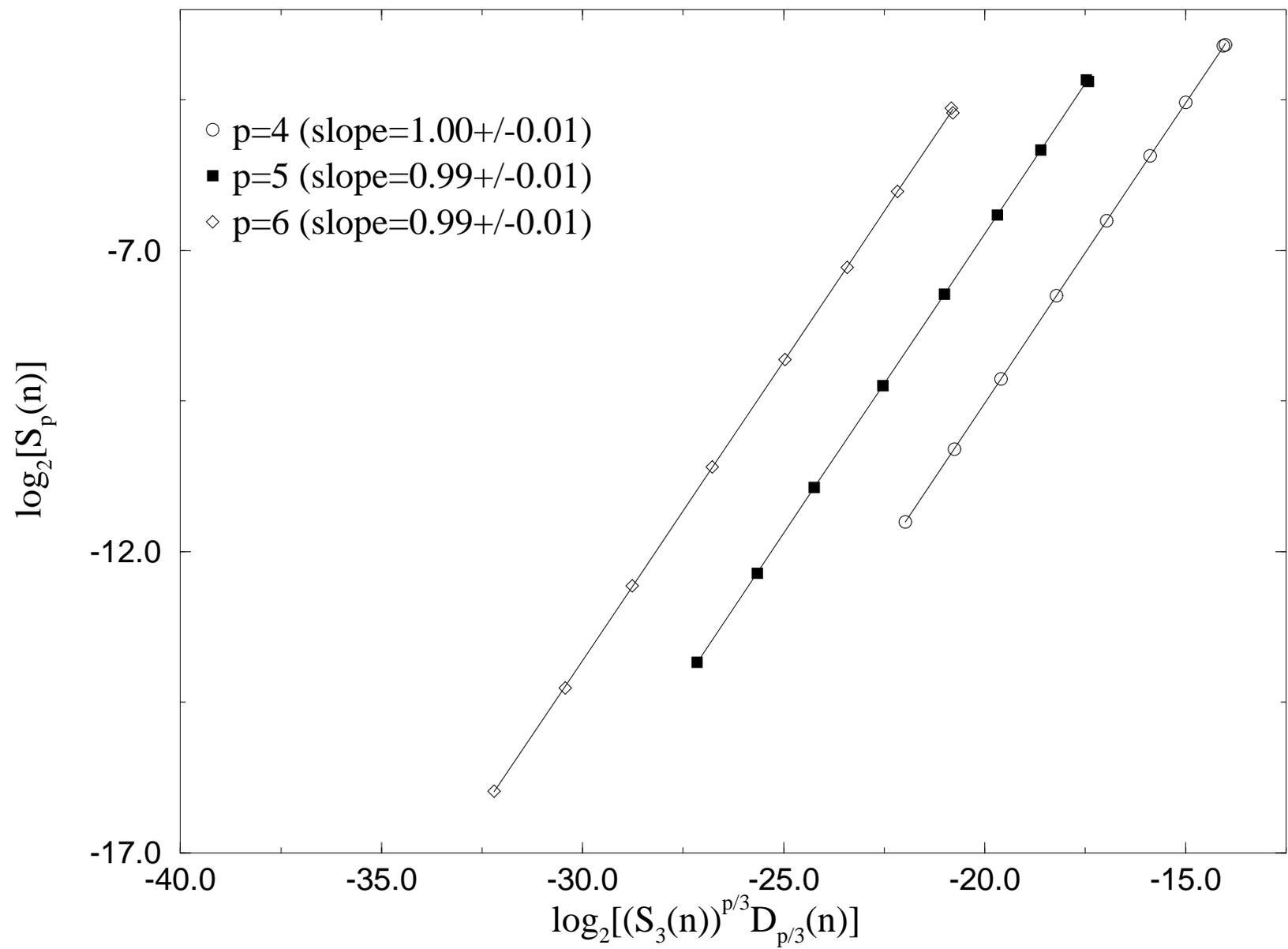
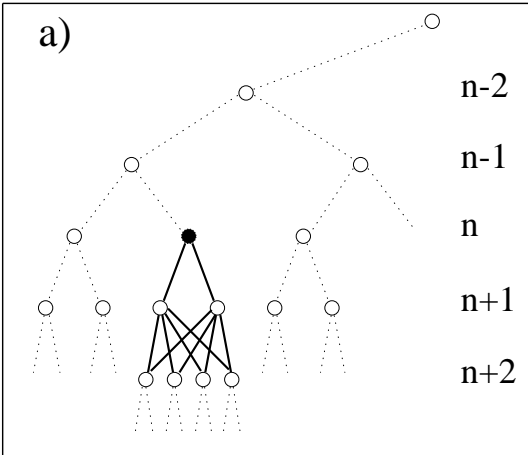
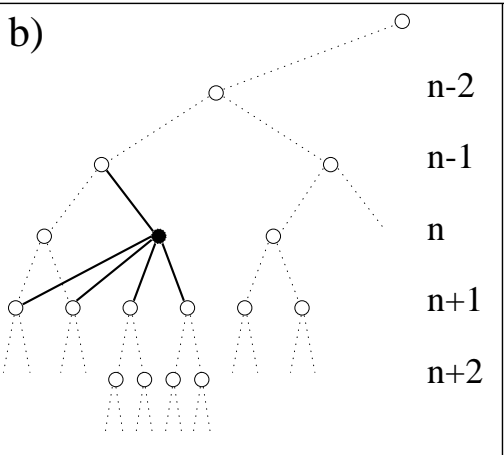
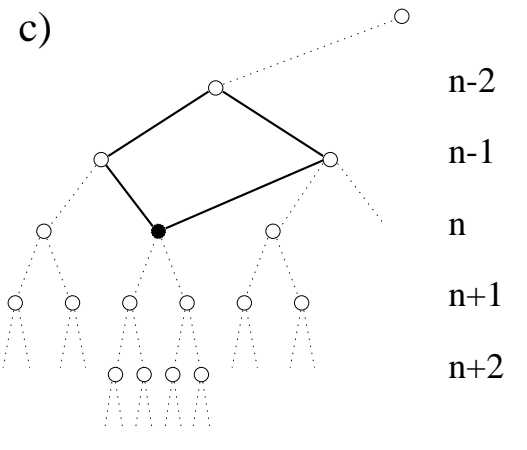
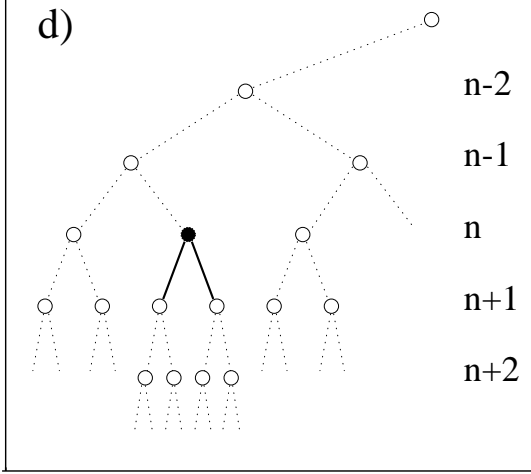
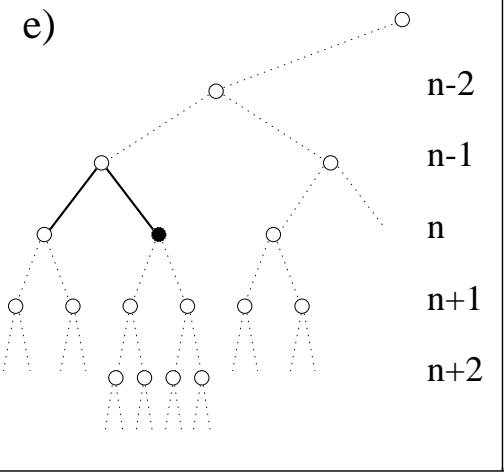
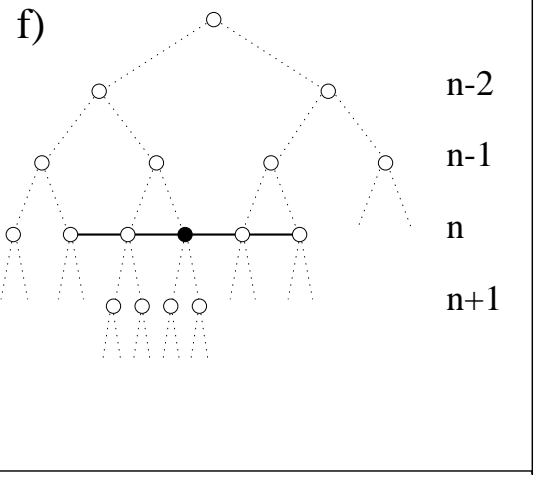
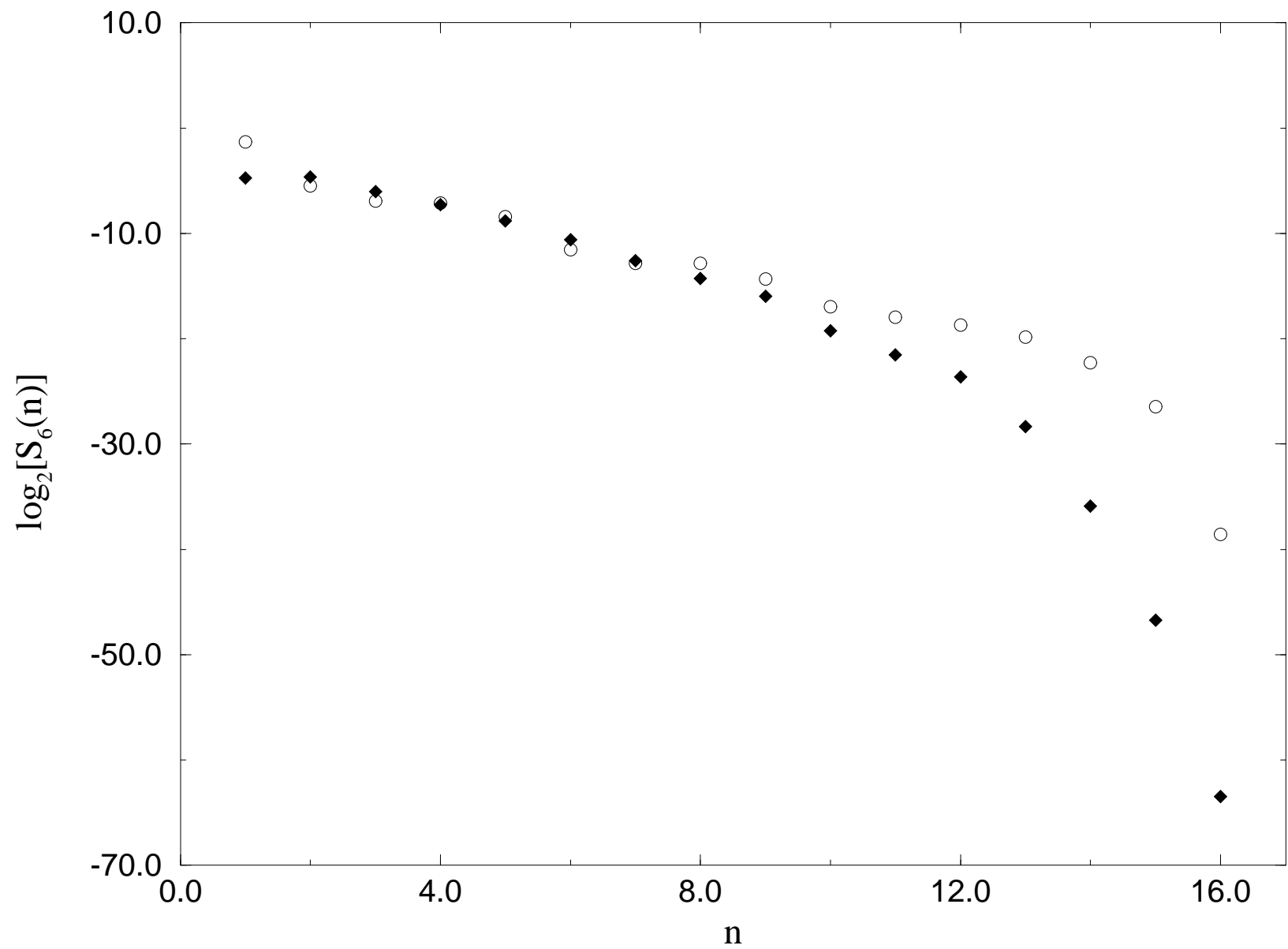
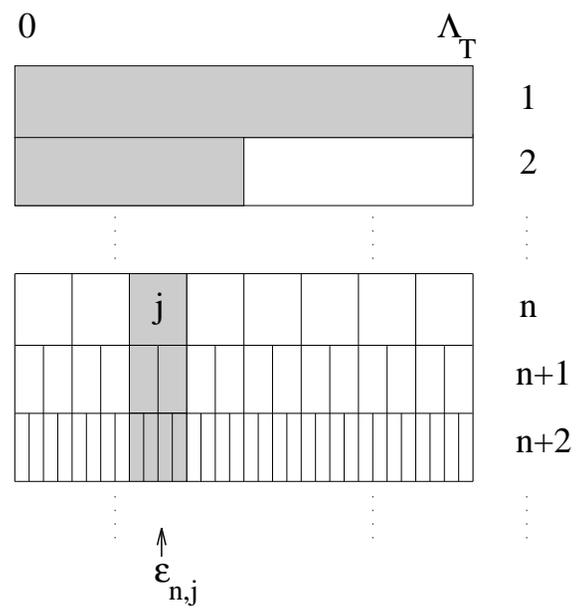
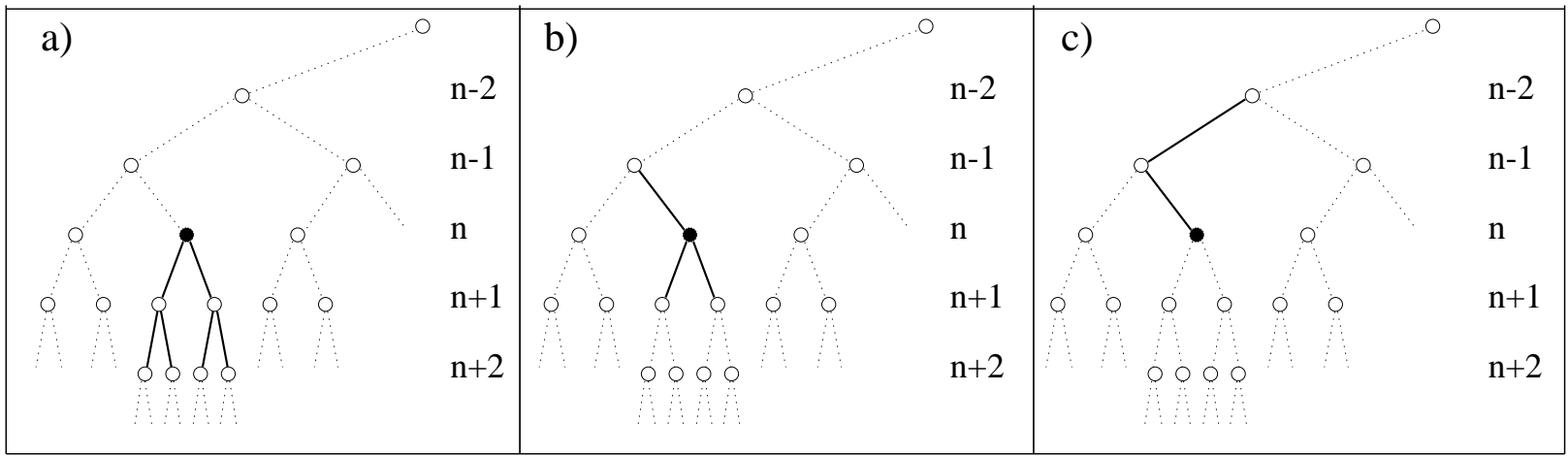
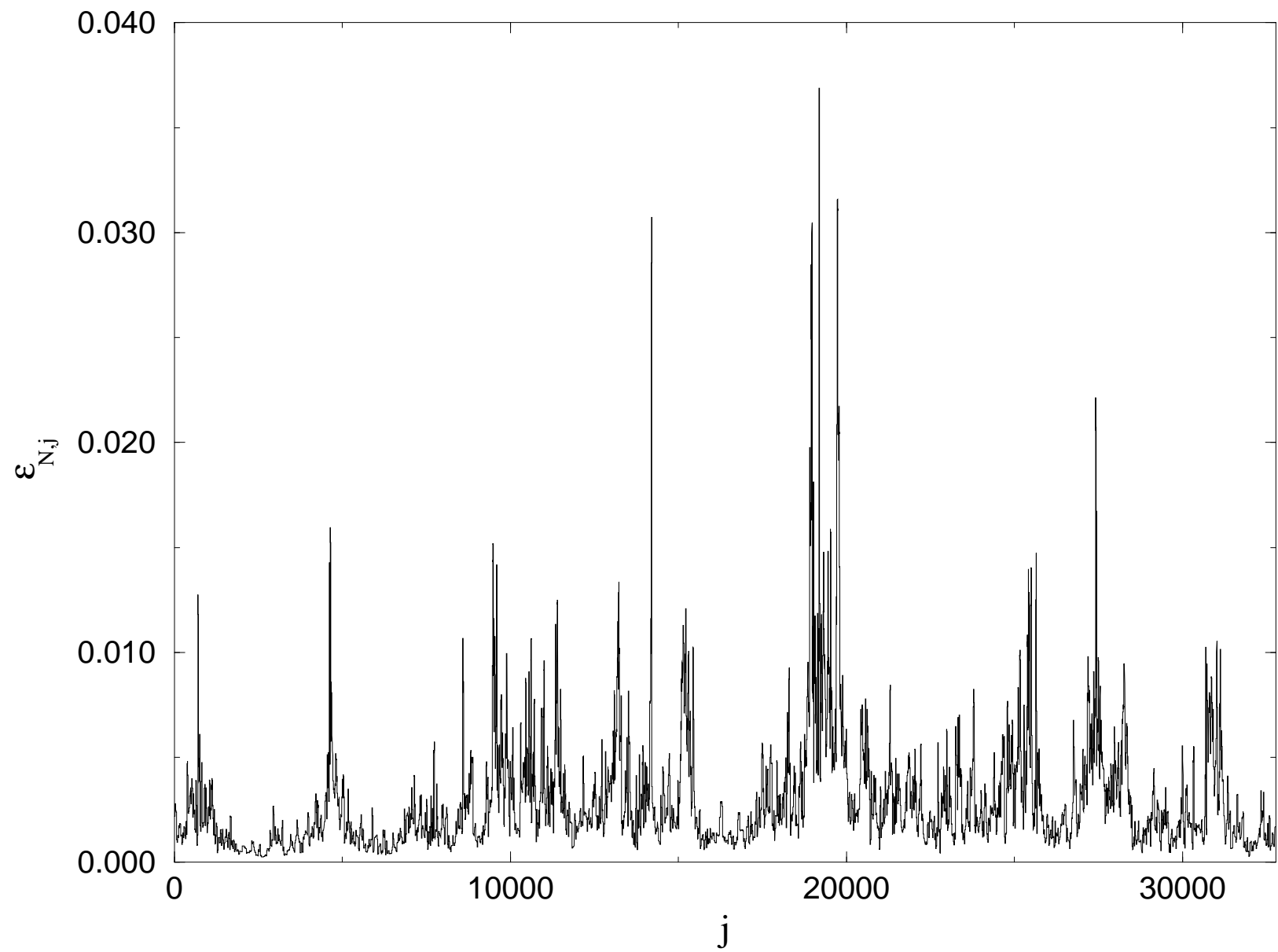
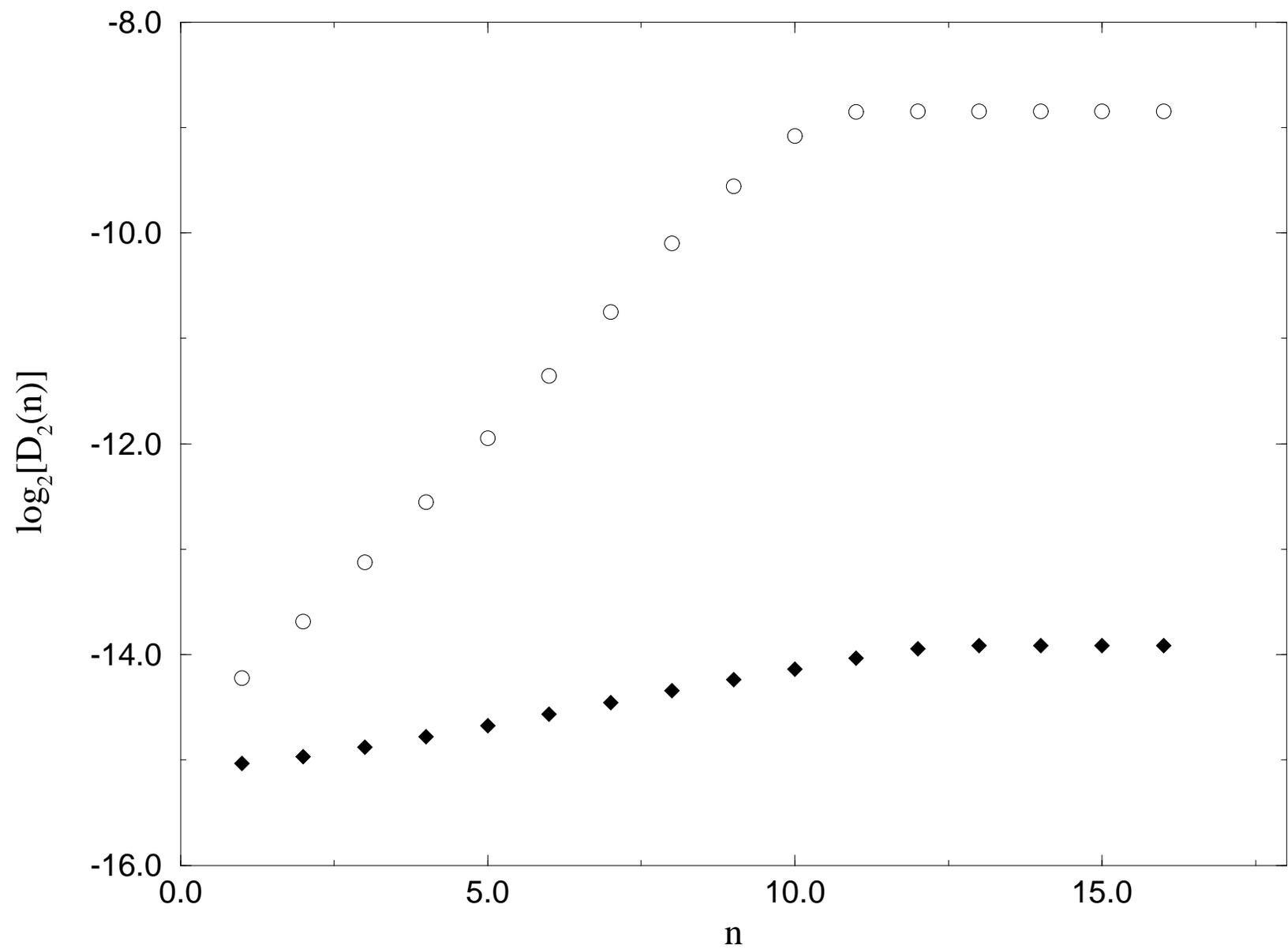
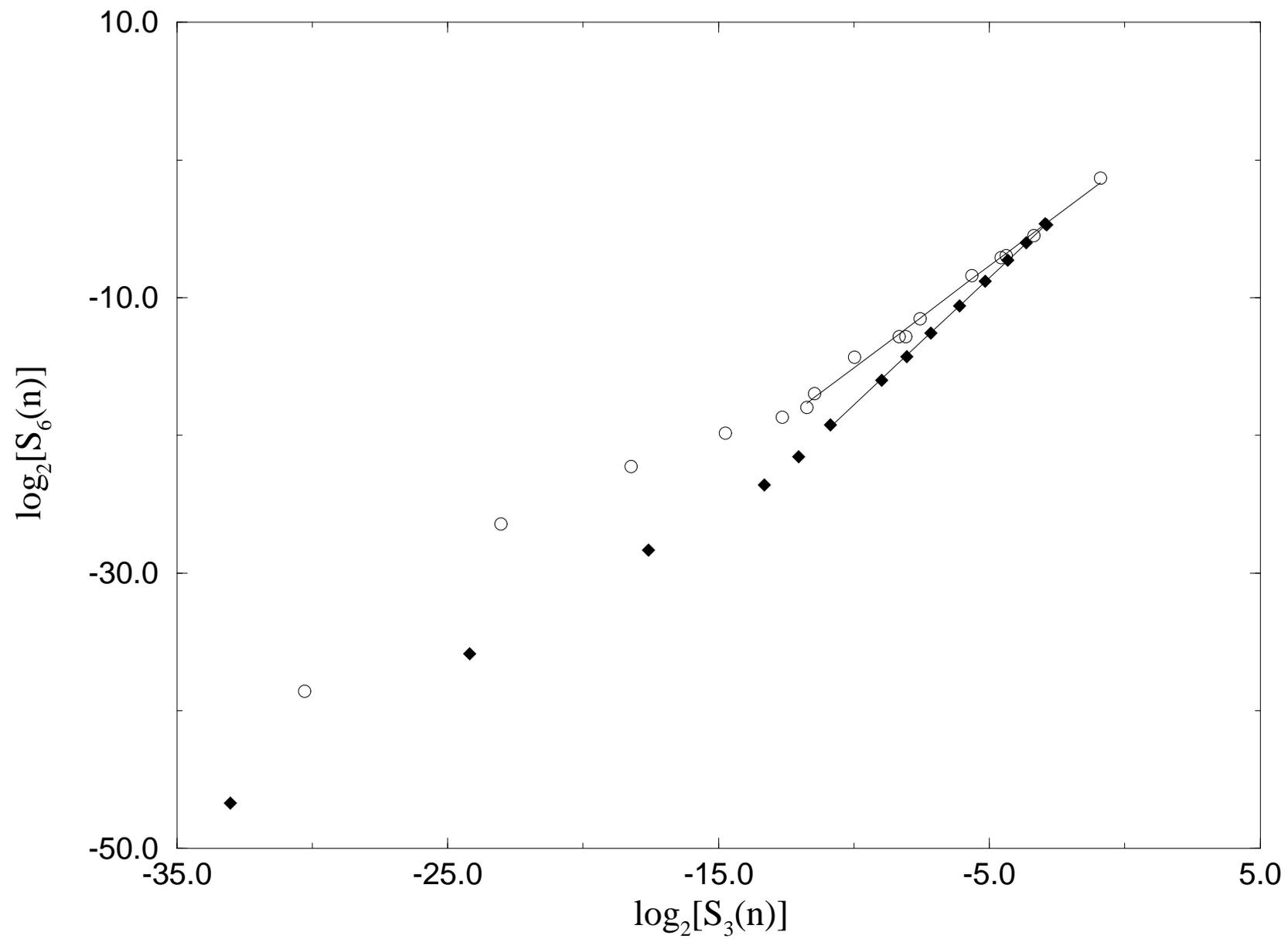
\end{document}